\documentclass[12pt]{article}
\DeclareUnicodeCharacter{2009}{\,}
\usepackage[utf8]{inputenc}
\usepackage{graphicx}
\usepackage{amsmath, amssymb}
\usepackage{hyperref}
\usepackage{geometry}
\usepackage{cite}
\usepackage{caption}
\usepackage{subcaption}
\usepackage{booktabs}
\usepackage{multirow}
\usepackage{float}
\usepackage{siunitx}

\begin{document}

\title{First-Principles Investigation of Sr\textsubscript{2}PrSbO\textsubscript{6} Double Perovskite: An Emerging Aspirant for Electrocatalysis, Plasmonic, Photonics, Thermoelectric and Solar Cell Applications}

\author{
    Md. Mohiuddin, Alamgir Kabir\textsuperscript{*}
}

\date{}

\maketitle

\begin{center}
    Department of Physics, University of Dhaka, Dhaka, Bangladesh. \\
    Email: \href{mailto:s-2019018090@phy.du.ac.bd}{s-2019018090@phy.du.ac.bd},
    \href{mailto:alamgir.kabir@du.ac.bd}{alamgir.kabir@du.ac.bd} \\
    \textsuperscript{*} Author to whom correspondence should be addressed
    
\end{center}

\begin{abstract}
In this study, we investigate the structural properties, chemical stability, electronic, optical, and thermoelectric properties of Sr\textsubscript{2}PrSbO\textsubscript{6} using first-principles calculations based on Density Functional Theory (DFT). The goal of this study is to evaluate its potential contribution to next-generation electrocatalysts, optoelectronic devices, and thermoelectric systems. The structural optimization reveals that Sr\textsubscript{2}PrSbO\textsubscript{6} crystallizes in a stable cubic perovskite structure with space group \textit{Fm-3m}. The calculated formation energy indicates high thermodynamic stability, confirming the viability of Sr\textsubscript{2}PrSbO\textsubscript{6} for practical applications. The electronic band structure calculations show that Sr\textsubscript{2}PrSbO\textsubscript{6} is a wide bandgap semiconductor with a direct bandgap of 3.488 eV at the $\Gamma$-point. The calculated density of states (DOS) indicates significant contributions from O 2\textit{p}, Sb 5\textit{p}, and Pr 5\textit{d} orbitals. Optical property calculations, including the dielectric function and absorption coefficient, reveal strong absorption in the UV regions, making Sr\textsubscript{2}PrSbO\textsubscript{6} a promising candidate for optoelectronic applications such as UV light-emitting diodes (LEDs) and photovoltaic-thermoelectric (PV-TE) tandem systems. At room temperature, the calculated dimensionless quantity ZT is 0.33, which indicates this material as a possible candidate for thermoelectric applications. Our results will serve as a benchmark for future experimental and theoretical research on the properties of this material.


\end{abstract}

\section{Introduction}

The increasing global demand for energy, in conjunction with the increasing apprehensions regarding the greenhouse effect and environmental pollution resulting from fossil fuel production, has prompted extensive research into the development of new electrochemical storage systems and renewable energy solutions.\cite{Walter2010,Zhang2015,Chen2011,Armand2008,Electrochemical1972,Kudo2008,Gasteiger2009,Bruce2012,Zhu2017,Boettcher2011} It is only recently that the creation of new functional materials, with unique functionality relevant to the described issues and invested to mitigate them, received intensive effort. Despite the progress, materials with unique and tailored properties are still essential for energy conversion processes, which vary across technologies. Quite impressively, the oxide perovskites and their derivatives\cite{Weng2017,Bandgap2018,Weng2018,Perez2018,Grinberg2013,Ge2018} are implemented within a range of technologies and shown to be versatile. Additionally, the structural and compositional flexibility of oxide perovskites provides properties like superconductivity, ferroelectricity, magnetoresistance, and ionic conductivity\cite{Pena2001,Cohen1992}.

In double perovskites, either the A-site or B-site is occupied by two different cations, leading to a general formula of the structure A$'$A$''$B\textsubscript{2}O\textsubscript{6} (double A-site) or A\textsubscript{2}B$'$B$''$O\textsubscript{6} (double B-site). Normally, A-site cations donate electrons to the [BO\textsubscript{6}] lattice, and B-site cations are the major contributors to the physical properties. Hence, double perovskites are named for the double B-site occupancy. These three-dimensional crystal structures are known for exhibiting the typical checkerboard geometry, in which the 'A' letters present rare-earth or large alkaline-earth elements; the 'B' and 'B$'$' present transition metal cations or lanthanides that showed multiferroic behavior in oxide systems and have garnered significant interest.\cite{Bandgap2018,García2012} The fact that the ionic radii or valence states of the B and B$'$ cations vary gives rise to spatial separation, which can encourage B-site ordering into a rock-salt structure, allowing room for electrostatic stability.\cite{García2012} It can then maintain the interactions of all charge, spin, and lattice systems in orchestration to allow the overall behavior of the material.Francisco Fernandez Martinez and co authors have synthesised Sr\textsubscript{2}RESbO\textsubscript{6} (RE=la to Lu and Y) and conducted FT-Raman and FT-IR vibrational spectroscopic studies.\cite{FernandezMartinez2012}

The distinctive 4\textit{f} electrons in rare earth elements are crucial for uncovering new properties and applications. Researchers are striving to bridge current gaps by providing insights into these materials' composition, synthesis, and uses.\cite{Chen2019} The double perovskites La\textsubscript{2}NiMnO\textsubscript{6} and Gd\textsubscript{2}NiMnO\textsubscript{6}, containing rare earth elements offer potential applications in thermoelectric, photocatalytic, magnetocaloric, and solar energy technologies.\cite{Rani2024} Ba\textsubscript{2}LnMO\textsubscript{6} double perovskites (where M = Sb, Bi, or Ln = lanthanides) were synthesized and their structures were examined in detail.\cite{Otsuka2015} Dielectric relaxation measurements were also conducted on a Pr\textsubscript{2}NiZrO\textsubscript{6} (PNZ) sample over a range of frequencies and temperatures spanning 313 K to 593 K.\cite{Mahato2016} Das et al. studied the electrical and magnetic properties of the rare earth double perovskite Dy\textsubscript{2}CoMnO\textsubscript{6}.\cite{Das2022} The perovskites A\textsubscript{2}FeMnO\textsubscript{6} (with A = Ba or La) were studied using both experimental approaches and Density Functional Theory to examine their structural, electronic, magnetic, and optical features.\cite{Dar2023} By employing both Density Functional Theory and experimental methods, the study explored the structural, electrical, and magnetic properties of RE\textsubscript{2}NiCrO\textsubscript{6} (RE = Ce, Pr, Nd).\cite{Kumar2022} Analyzed through DFT, the double perovskites Sr\textsubscript{2}(RE)CoO\textsubscript{6} showed promise for use in thermoelectric and photovoltaic solar technologies.\cite{Laghzaoui2023} A thorough first-principles study of Sr\textsubscript{2}ScBiO\textsubscript{6} revealed its promising physical properties for low-cost energy technologies.\cite{AlQaisi2023} Sr\textsubscript{2}XNbO\textsubscript{6} (where X = La or Lu) was examined in a separate research for its potential in UV thermoelectric and optoelectronic uses, revealing direct bandgaps of 4.02 eV and 3.7 eV.\cite{Hanif2022}In a similar vein, Haid and co-authors explored Sr\textsubscript{2}CrTaO\textsubscript{6}, underscoring its half-metallic ferrimagnetic ground state and its applications in visible, ultraviolet, and infrared technologies.\cite{Haid2019}

The literature on double perovskites with Pr at the B-site and Sb at the B$'$-site is noticeably lacking, despite the fact that oxide double perovskites have many applications.Sr\textsubscript{2}PrSbO\textsubscript{6} is one such material with a perovskite structure, making it a promising candidate for these applications.To our best knowledge  The cubic symmetry and high thermodynamic stability of Sr\textsubscript{2}PrSbO\textsubscript{6} suggest that it could be used in high-performance electronic devices\cite{Sun2017}.This gap represents a substantial opportunity to investigate these materials, as the distinctive compositional structure of Sr\textsubscript{2}PrSbO\textsubscript{6} double perovskites may yield remarkable electronic properties and varied potential applications. Consequently, this study seeks to examine the versatility of Sr\textsubscript{2}PrSbO\textsubscript{6}  in advanced technologies, including solid oxide fuel cells, catalysis, optoelectronics, and thermoelectrics, utilizing Density Functional Theory (DFT).

Recent advancements in Density Functional Theory (DFT) have enabled detailed insights into the structural, electronic, and optical properties of complex oxides like Sr\textsubscript{2}PrSbO\textsubscript{6}. To the best of our knowledge, no previous studies have extensively reported the structural, optoelectronic, thermoelectric properties, and chemical stability of this material. Hence, by employing first-principles calculations, this study aims to provide a comprehensive understanding of Sr\textsubscript{2}PrSbO\textsubscript{6}, including its electronic band structure, density of states, optical properties (such as the dielectric function and absorption coefficient), and chemical stability. These properties are crucial for evaluating the material's potential for use in devices such as UV light-emitting diodes (LEDs), photovoltaic cells, and high-power electronics. \cite{Hudgins2003}

\section{Computational Methodology}

In this study, first-principles calculations based on Density Functional Theory (DFT) were utilized to examine the structural, electronic, optical properties, and chemical stability of Sr\textsubscript{2}PrSbO\textsubscript{6}. These calculations were carried out using the Vienna Ab initio Simulation Package (VASP)\cite{Hohenberg1964,Payne1992,Kohn1965,Gillan1989,Vanderbilt1990,Kresse1996}.Augmented wavefunctions, such as Projector Augmented Wave (PAW) potentials, enhance the fidelity of density functional theory (DFT) calculations. They do so by accurately representing electron density near atomic cores and improving the modeling of electron-core interactions.\cite{Kresse1999}. The utilization of PAW potentials ensures a more accurate depiction of electronic properties while effectively considering core electron interactions. Geometrical optimizations are conducted using the Perdew-Burke-Ernzerhof (PBE) functional based on the Generalized Gradient Approximation (GGA) for exchange-correlation interactions\cite{Paier2005,Perdew1996}.The GGA-PBE functional was chosen because it is computationally efficient, and has significantly higher accuracy than the LDA functional across a variety of systems. Ions are relaxed into their lowest energy state configurations using the conjugate-gradient technique.   The electron-ion interactions were treated with Projector Augmented Wave (PAW) pseudopotentials, specifically chosen from the standard VASP library to match the PBE functional for elements Sr, Pr, Sb, and O. This selection ensures a reliable representation of core-electron interactions while maintaining computational efficiency.

A plane-wave basis set with a kinetic energy cutoff of 400 eV was applied, following thorough convergence tests to validate the energy cutoff and k-point mesh for accuracy and reliability. The Brillouin zone was sampled using  Monkhorst-Pack\cite{Wang2021} grid of $3 \times 3 \times 3$ for both structural optimization and electronic property calculations, confirmed sufficient through convergence testing.  The initial structure of Sr\textsubscript{2}PrSbO\textsubscript{6} was sourced from the Materials Project database, and subsequent structural optimization was conducted using the Conjugate Gradient (CG) algorithm with specific convergence criteria: a total energy convergence of $1 \times 10^{-8}$ eV/atom, a force convergence threshold of $-2 \times 10^{-2}$ eV/Å, and a fully relaxed stress tensor and cell shape.

Electronic band structure calculations were performed along the high-symmetry path in the Brillouin zone, specifically the $\Gamma$-X-U$|$K-$\Gamma$-L-W-X points, along with the computation of the density of states (DOS) and partial density of states (PDOS) to analyze orbital contributions. A spin-polarized DFT approach was employed for band structure, with a non-self-consistent calculation initiated from existing wavefunctions and charge densities from a prior self-consistent field (SCF) calculation.  Similarly, DOS and PDOS calculations used spin-polarized DFT.The projected density of states (PDOS) was enabled for detailed orbital analysis.

Optical properties, including the dielectric function, refractive index, and absorption coefficient, were derived using Kohn-Sham orbitals from the DFT calculations. The frequency-dependent dielectric function, $\varepsilon(\omega) = \varepsilon_1(\omega) + i\varepsilon_2(\omega)$, was evaluated to understand the interaction of Sr\textsubscript{2}PrSbO\textsubscript{6} with electromagnetic waves.  Chemical stability was analyzed through formation energy calculations.The transport coefficients were
determined using the semiclassical Boltzmann theory as implemented in the BoltzTraP2 code\cite{Madsen2018}.
Overall, these comprehensive calculations provide valuable insights into the fundamental properties of Sr\textsubscript{2}PrSbO\textsubscript{6}, with a focus on accurate structural, electronic, and optical characteristics supported by thorough convergence testing and methodological rigor.

\section{Structural Properties}

\subsection{Crystal Structure}
\begin{figure}[H]
    \centering
    \includegraphics[width=1\linewidth]{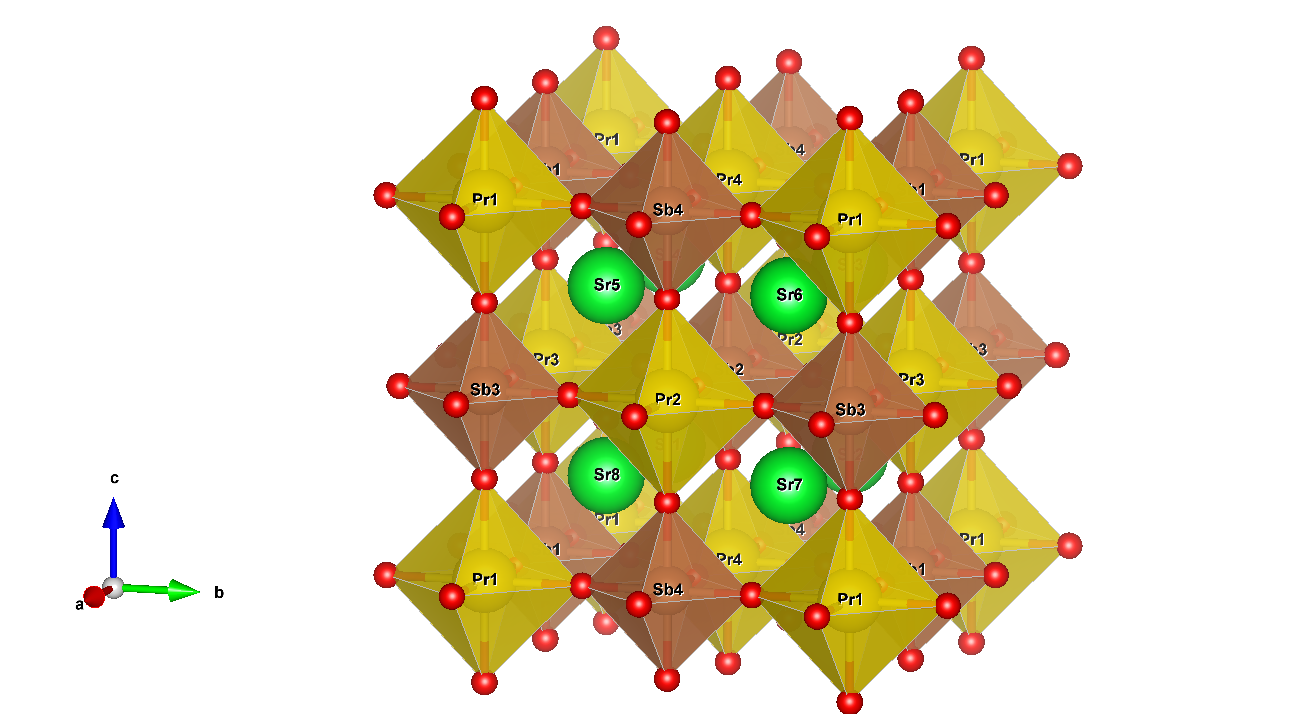}
    
           \caption{Crystal structure of the compound showing the arrangement of Pr (yellow polyhedra), Sb (brown polyhedra), and Sr (green spheres) atoms. The coordination geometry is depicted with oxygen atoms (red spheres), highlighting the connectivity of the polyhedra. The unit cell directions (\textit{a}, \textit{b}, and \textit{c}) are shown in the bottom-left corner for reference.}

    \label{fig:structure}
\end{figure}

Sr\textsubscript{2}PrSbO\textsubscript{6} crystallizes in a cubic crystal system with a space group of \textit{Fm-3m} which can be observed from Figure~\ref{fig:structure}. The optimized lattice parameters are $a = 4.267$ Å, $b = 4.267$ Å, and $c = 4.267$ Å, reflecting the high degree of symmetry inherent in a cubic structure. The equivalence of the lattice parameters confirms the ideal cubic symmetry of the crystal. The unit cell consists of 2 formula units of Sr\textsubscript{2}PrSbO\textsubscript{6}.

In this structure, Sr atoms occupy the corner positions with coordinates at (0.25, 0.25, 0.25) and (0.75, 0.75, 0.75). Pr atoms are located at the body center with coordinates at (0.0, 0.0, 0.0). Sb atoms occupy the face-centered positions with coordinates at (0.5, 0.5, 0.5). O atoms are positioned at the general positions with the following coordinates: (0.7323, 0.2677, 0.2677), (0.2677, 0.7323, 0.7323), (0.2677, 0.7323, 0.2677), (0.7323, 0.2677, 0.7323), (0.2677, 0.2677, 0.7323), and (0.7323, 0.7323, 0.2677). The detailed atomic positions and corresponding Wyckoff coordinates are presented in Table~\ref{table:atomic_positions}.

\begin{table}[H]
\centering
\caption{Atomic positions and Wyckoff coordinates for Sr\textsubscript{2}PrSbO\textsubscript{6}.}
\label{table:atomic_positions}
\begin{tabular}{lllll}
\toprule
Atom & Wyckoff & x      & y      & z      \\
\midrule
Sr   & 8c      & 0.25   & 0.25   & 0.25   \\
Pr   & 4a      & 0.0    & 0.0    & 0.0    \\
Sb   & 4b      & 0.5    & 0.5    & 0.5    \\
O    & 24e     & 0.7323 & 0.2677 & 0.2677 \\
O    & 24e     & 0.2677 & 0.7323 & 0.7323 \\
O    & 24e     & 0.2677 & 0.7323 & 0.2677 \\
O    & 24e     & 0.7323 & 0.2677 & 0.7323 \\
O    & 24e     & 0.2677 & 0.2677 & 0.7323 \\
O    & 24e     & 0.7323 & 0.7323 & 0.2677 \\
\bottomrule
\end{tabular}
\end{table}

\subsection{Bond Lengths and Angles}

The bond lengths and angles in the Sr\textsubscript{2}PrSbO\textsubscript{6} structure were analyzed in detail to understand the nature of bonding and the geometry within the crystal lattice. The Sr–O bond length was measured at 3.02109 Å, indicating a weaker ionic interaction due to the larger ionic radius of Sr$^{2+}$. In contrast, the Pr–O and Sb–O bond lengths were found to be 2.28474 Å and 1.98238 Å, respectively, with the shorter Sb–O bond length suggesting a more covalent character due to the higher electronegativity of Sb. The bond angle analysis revealed an O–Sb–O angle of 90.0000 degrees, which is typical for a nearly perfect octahedral coordination, implying minimal distortion in the SbO\textsubscript{6} octahedra. Additionally, the Sr–O–Pr bond angle was calculated to be 87.1316 degrees, indicating a slight deviation from an ideal linear arrangement, likely resulting from the size mismatch between Sr, Pr, and Sb atoms, leading to minor structural distortions.The result is summarized in Table~\ref{table:bond_lengths_angles}.

\begin{table}[H]
\centering
\caption{Bond lengths and angles in SrPrSbO\textsubscript{6}.}
\label{table:bond_lengths_angles}
\begin{tabular}{ll}
\toprule
Bond/Angle & Value \\
\midrule
Sr--O Bond Length & 3.02109 Å \\
Pr--O Bond Length & 2.28474 Å \\
Sb--O Bond Length & 1.98238 Å \\
O--Sb--O Angle & 90.0000 degrees \\
Sr--O--Pr Angle & 87.1316 degrees \\
\bottomrule
\end{tabular}
\end{table}

\subsection{Tolerance Factor}
To understand the structural stability and the formability of the perovskite crystal structures, the Goldschmidt tolerance factor \cite{Sato2016} that compares the cation's and anion's ionic sizes in a crystal structure, is studied. The Sr\textsubscript{2}PrSbO\textsubscript{6} stoichiometry defines the overall crystallographic structure of double-perovskite material, where  A  represents Sr,  B  being the lanthanide series metal cation Pr,  B$'$  refers to the post-transition metal cation Sb, and \( X \) is oxygen. For Sr\textsubscript{2}PrSbO\textsubscript{6}, the tolerance factor (\( T_f \)) is defined by the ionic radii of the \( A \) (\( R_A \)), \( B \) (\( R_B \)), \( B' \) (\( R_B' \)) and \( O \) (\( R_O \)) site ions as stated in equation (\ref{eq1}).

\begin{equation}
T_f = \frac{R_A + R_O}{\sqrt{2} \left[ \frac{R_B + R_{B'}}{2} + R_O \right]}
\label{eq1}
\end{equation}

For an ideal cubic perovskite structure, \( T_f \) typically ranges between 0.8 and 1.0, with the  A -cation being larger than the  B-cation \cite{Jin2021}. However, when the  A -cation is smaller, \( T_f \) decreases below 1.0, leading to octahedral tilting to fill the space, which in turn reduces the symmetry of the crystal structure. 

The calculated tolerance factor of 0.968 is within the stable range for perovskite structures ($0.8 \leq t \leq 1.0$), indicating that Sr\textsubscript{2}PrSbO\textsubscript{6} is structurally stable and likely to maintain its perovskite phase under standard conditions.\cite{Cohen1992,Sato2016} 

\subsection{Thermodynamic Stability}

The thermodynamic stability of Sr\textsubscript{2}PrSbO\textsubscript{6} was evaluated through the calculation of its formation energy using the equation $\Delta E_{\text{formation}} = \sum E_{\text{products}} - \sum E_{\text{reactants}}
$.\cite{Melius1992} The formation energy for Sr\textsubscript{2}PrSbO\textsubscript{6} was found to be $-22.7698127$ eV per formula unit, which corresponds to $-2.27698127$ eV per atom. This significantly negative value indicates high thermodynamic stability, confirming that Sr\textsubscript{2}PrSbO\textsubscript{6} is stable and can be synthesized under appropriate conditions. The negative formation energy further affirms that Sr\textsubscript{2}PrSbO\textsubscript{6} is not prone to decomposition, making it a promising candidate for various applications in the perovskite structure.\cite{García2012}
\section{Chemical Stability}

\begin{figure}[H]
    \centering

                        \centering
                        \includegraphics[width=1\linewidth]{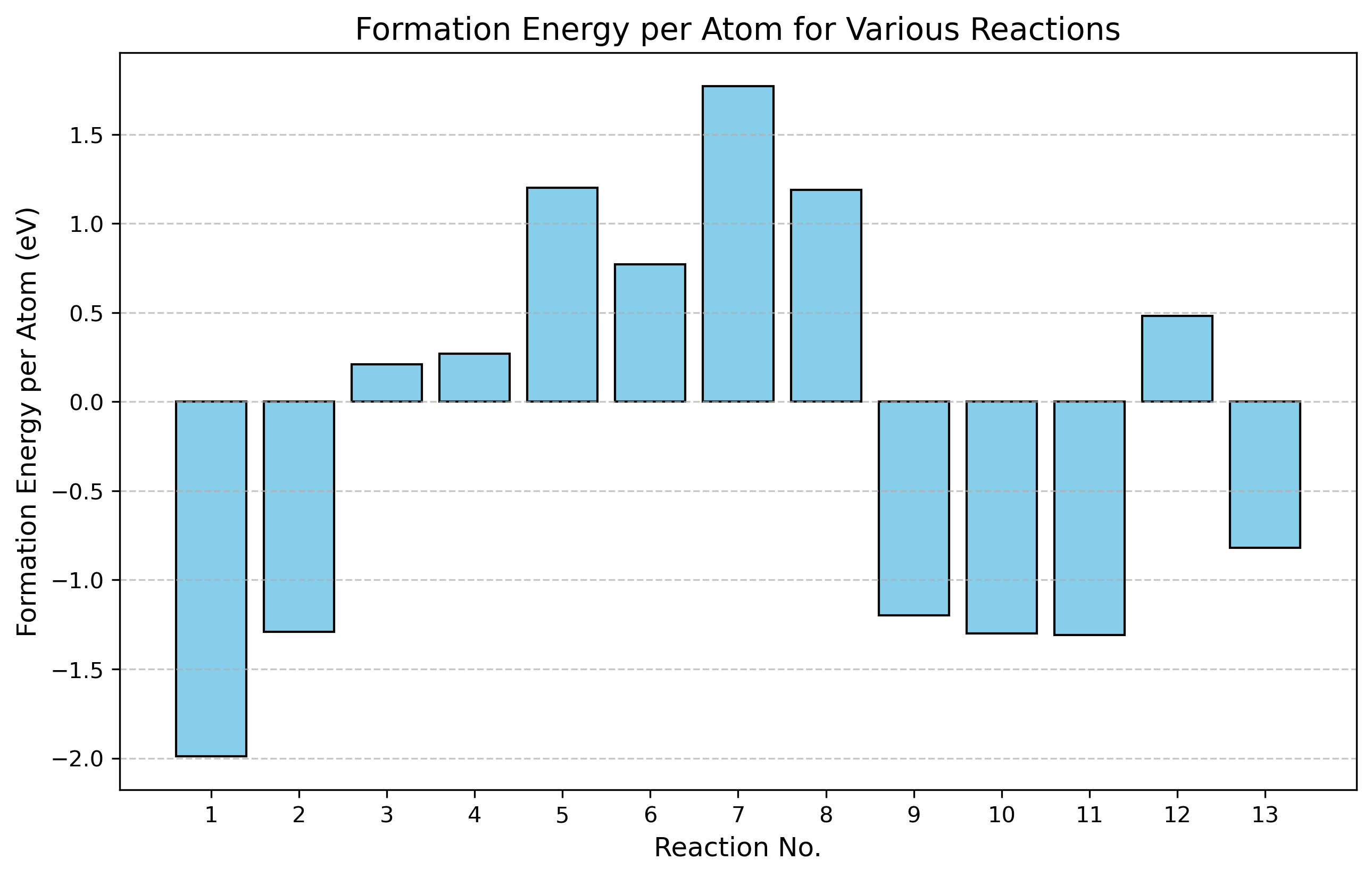}

        \caption{Formation energy per atom (eV) for various decomposition reactions of Sr\textsubscript{2}PrSbO\textsubscript{6}, labeled by reaction number. Positive formation energy values indicate energetically unfavorable reactions, while negative values correspond to energetically favorable reactions. Reaction 7 exhibits the highest positive formation energy, indicating a strong endothermic process, whereas reactions 1 and 2 display the lowest formation energies, suggesting highly exothermic reactions.}

    \label{fig:chemical}
\end{figure}

The formation energy of the decomposed compound is summarized in Table~\ref{table:formation_energies} and Figure~\ref{fig:chemical}.From the analysis of the decomposition reactions of Sr\textsubscript{2}PrSbO\textsubscript{6}, based on the calculated formation energies per atom, the following conclusions can be made:

The most stable decomposition pathway for Sr\textsubscript{2}PrSbO\textsubscript{6} is represented by Reaction 1 ($2\text{SrO} + \text{Pr}_2\text{O}_3 + \text{Sb}_2\text{O}_3$) and Reaction 9 ($\text{Sr}_2\text{O}_3 + \text{Pr}_2\text{O}_3 + \text{Sb}_2\text{O}_3$), having the most negative formation energies per atom of $-1.99$ eV and $-1.97$ eV, respectively. This suggests that Sr\textsubscript{2}PrSbO\textsubscript{6} is most likely to decompose into SrO, Pr\textsubscript{2}O\textsubscript{3}, and Sb\textsubscript{2}O\textsubscript{3}, or alternatively into Sr\textsubscript{2}O\textsubscript{3}, Pr\textsubscript{2}O\textsubscript{3}, and Sb\textsubscript{2}O\textsubscript{3}, indicating a relative stability under standard conditions but with potential for decomposition under specific scenarios.

Reactions with positive formation energies, such as Reaction 3 ($2\text{SrO} + \text{PrSbO}_4$, 0.02 eV) and Reaction 12 ($\text{SrO} + \text{SrPrO}_4 + \text{Sb}$, 0.48 eV), are less favorable, indicating that these products are less likely to form spontaneously. Intermediate stability is observed in reactions like Reaction 5 ($2\text{SrO} + \text{PrO}_3 + \text{Sb}_2\text{O}_5$, $-1.70$ eV) and Reaction 10 ($\text{Sr}_2\text{O}_3 + \text{PrSbO}_4 + \text{SbO}_3$, $-1.30$ eV), suggesting partial decomposition under specific environmental conditions such as elevated temperatures.

Unstable reactions, such as Reaction 6 ($2\text{Sr} + \text{Pr} + \text{Sb} + 3 \text{O}_2$, 2.28 eV) and Reaction 7 ($2\text{SrO} + \text{Pr} + \text{Sb} + 1.5 \text{O}_2$, 1.77 eV), indicate that decomposition into elemental forms is highly unlikely. Overall, Sr\textsubscript{2}PrSbO\textsubscript{6} demonstrates considerable stability, making it suitable for high-temperature and chemically resistant applications, such as in solid oxide fuel cells, catalysis, and dielectric materials, reinforcing its potential use in demanding environments.
\begin{table}[H]
\centering
\caption{Formation energies per atom for the decomposition reactions of $\mathrm{Sr_2PrSbO_6}$.}
\label{table:formation_energies}
\begin{tabular}{cccc}
\toprule
Reaction No. & Full Reaction & Total Atoms & Formation Energy per Atom (eV) \\
\midrule
1  & $2\mathrm{SrO} + \mathrm{Pr_2O_3} + \mathrm{Sb_2O_3}$                     & 14 & $-1.99$ \\
2  & $2\mathrm{SrO} + \mathrm{Pr_2O_3} + 2\mathrm{Sb}$                         & 11 & $-1.29$ \\
3  & $2\mathrm{SrO} + \mathrm{Pr_2O_3} + \mathrm{SbO_2}$                        & 14 & $-1.20$ \\
4  & $2\mathrm{SrO} + \mathrm{Pr_2O_3} + \mathrm{SbO_4}$                        & 10 & $0.02$ \\
5  & $2\mathrm{SrO} + \mathrm{PrO_2} + \mathrm{Sb_2O_3}$                        & 14 & $0.21$ \\
6  & $2\mathrm{SrO} + \mathrm{PrO_2} + \mathrm{SbO_2}$                          & 15 & $-1.70$ \\
7  & $2\mathrm{Sr} + \mathrm{Pr} + \mathrm{Sb} + 3\mathrm{O_2}$                 & 10 & $2.28$ \\
8  & $2\mathrm{SrO} + \mathrm{Pr} + \mathrm{Sb} + \tfrac{1}{2}\mathrm{O_2}$     & 12 & $0.07$ \\
9  & $2\mathrm{SrO} + \mathrm{PrO_3} + \mathrm{Sb} + \mathrm{SbO_2}$            & 13 & $-1.97$ \\
10 & $\mathrm{Sr_2O_3} + \mathrm{PrSbO_4} + \mathrm{SbO_3}$                     & 15 & $-1.30$ \\
11 & $2\mathrm{SrO} + \mathrm{PrO_2} + \mathrm{SbO_2}$                          & 14 & $-0.98$ \\
12 & $\mathrm{SrO} + \mathrm{SrPrO_4} + \mathrm{Sb}$                            & 9  & $0.48$ \\
13 & $\mathrm{SrO} + \mathrm{SrPrO_3} + \mathrm{Sb_2O_3}$                       & 11 & $-0.82$ \\
\bottomrule
\end{tabular}
\end{table}

\section{Electronic Properties}

\subsection{Band Structure}

Electronic properties offer crucial insights into various material characteristics, such as the nature of bonding between anions and cations, photon absorption efficiency, electrical conductivity, and related attributes. Recently, lanthanide-based double perovskites have attracted attention for their unique direct band gap, making them a subject of significant research interest.\cite{Volonakis2016,Charifi2023}Additionally, the potential of these materials as emitters of white light has been investigated in pioneering study regarding their band gap characteristics, the experimental synthesis of LD has produced favorable findings.\cite{Abdullah2023}.

\begin{figure}[H]
    \centering

            \includegraphics[width=1\linewidth]{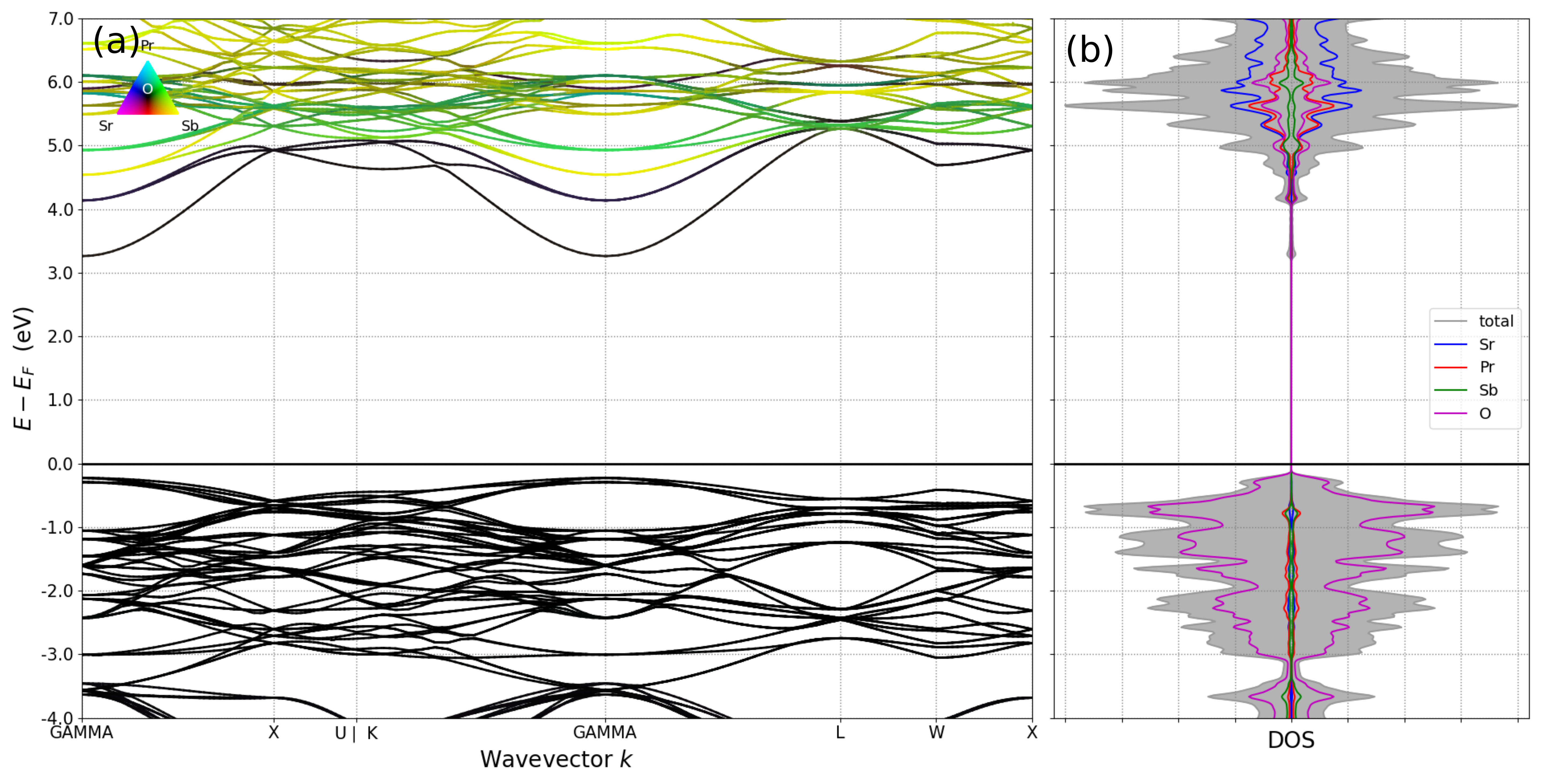}

       \caption{Spin-polarized electronic properties of Sr\textsubscript{2}PrSbO\textsubscript{6} determined using GGA calculations: (a) Electronic band structure along high-symmetry points in the Brillouin zone, with the Fermi level set to zero; (b) Total and partial density of states (DOS and PDOS), showing contributions from Sr, Pr, Sb, and O atoms. The inset highlights the contributions from individual atomic orbitals.}

    \label{fig:bandstructure}
\end{figure}

The electronic band structure of Sr\textsubscript{2}PrSbO\textsubscript{6} was calculated using Density Functional Theory (DFT) along the high-symmetry paths in the Brillouin zone, specifically $\Gamma$-X-U$|$K-$\Gamma$-L-W-X. The calculated band structure is shown in Figure~\ref{fig:bandstructure}(a).As seen in the figure, Sr\textsubscript{2}PrSbO\textsubscript{6} exhibits a direct bandgap of 3.488 eV, located at the $\Gamma$-point for both spin-up and spin-down configurations. This significant bandgap suggests that Sr\textsubscript{2}PrSbO\textsubscript{6} is a wide bandgap semiconductor, which makes it suitable for applications in optoelectronic devices such as ultraviolet (UV) light emitters, and potential applications in high-power electronics.\cite{Hudgins2003}Additionally, its large bandgap could imply strong resistance to thermal excitation, suggesting possible applications in high-temperature electronic devices.\cite{Touati2006}

\subsection{Density of States}
The total and partial density of states (TDOS and PDOS) are shown in Figure~\ref{fig:bandstructure}(b). The TDOS also confirms that the band gap is 3.488 eV.This  wide bandgap makes Sr\textsubscript{2}PrSbO\textsubscript{6} suitable for tandem solar cell.\cite{Chen2024,Forgacs2017,Fang2024}  
The total and partial density of states (TDOS and PDOS) analyses of Sr\textsubscript{2}PrSbO\textsubscript{6} indicate that it is a non-magnetic semiconductor with a bandgap at the Fermi level, where the symmetric spin-up and spin-down states confirm the absence of spin polarization. The valence band is predominantly composed of O 2\textit{p} states with minor contributions from Sb 5\textit{p} and Pr 4\textit{f} states, highlighting the crucial role of oxygen in the electronic structure. In the conduction band, Sb 5\textit{d} states are dominant, with some involvement of Pr 5\textit{p} states, while the presence of Pr 4\textit{f} states near the Fermi level suggests potential optical interactions. This electronic configuration makes Sr\textsubscript{2}PrSbO\textsubscript{6} possible candidate for several applications, including optoelectronics (e.g., UV light-emitting diodes and laser diodes)\cite{Shahzad2024}, photodetectors operating in the UV-visible spectrum\cite{Zhou2023} due to its wide bandgap and ability to perform under higher voltages and temperatures. 
Spin-polarized calculations were performed to investigate the magnetic properties of Sr\textsubscript{2}PrSbO\textsubscript{6}. The results indicate that Sr\textsubscript{2}PrSbO\textsubscript{6} is non-magnetic, with no net magnetic moment per unit cell. The absence of magnetism is primarily attributed to the electronic configuration and the lack of significant spin polarisation in the Pr atoms. The calculated spin density shows negligible contributions from all atoms within the unit cell, confirming the non-magnetic nature of the compound.
The high activity and stability of  double perovskites can be explained by having the O p-band centre neither too close nor
too far from the Fermi level.In addition to electronic structure, the covalency of the B–O bond also plays a critical role in oxygen reduction reaction(ORR). A stronger covalency of the
B–O bond should increase the driving force and thereby facilitate the O2–/OH– exchange on the
surface B ions, which can be considered as the rate-limiting step of ORR\cite{Grimaud2013}.The O p-band centre of our material falls under this criteria, which makes it a possible candidate for electrocatalysis applications. 
\section{Optical Properties }

\begin{figure}[H]
    \centering
    
        \includegraphics[width=1\linewidth]{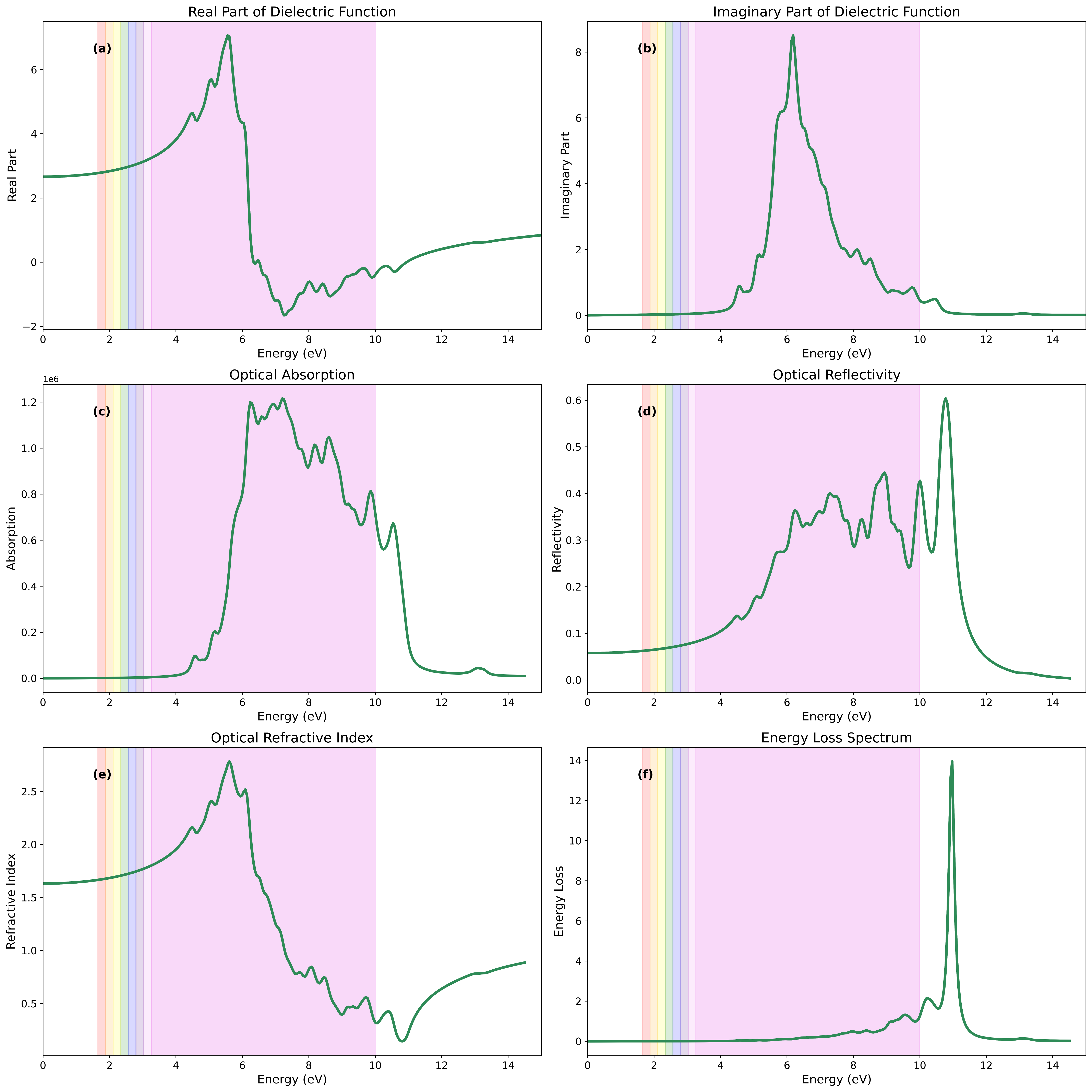}

        \caption{Optical properties of Sr\textsubscript{2}PrSbO\textsubscript{6} as a function of photon energy: (a) Real part of the dielectric function; (b) Imaginary part of the dielectric function; (c) Optical absorption spectrum; (d) Optical reflectivity; (e) Refractive index; (f) Electron energy loss spectrum (EELS). Highlighted regions indicate the energy ranges of interest.}

    \label{fig:optical}
\end{figure}

The optical properties codify key insights into aspects such as the electronic energy band and phonon vibrational modes \cite{Fox2010, SpringerLink2022}. We have examined the optical characteristics of Sr$_2$PrSbO$_6$ double perovskite in the energy range of $0 - 15\,\text{eV}$ in order to investigate their potential photonic and optoelectronic uses. The Kramers-Kronig relation for the complex dielectric function\cite{Albanesi2005,Kumar2020}
\begin{equation}
\varepsilon(\omega) = \varepsilon_1(\omega) + i\varepsilon_2(\omega)
\end{equation}
can be used to express the dielectric properties. In this relation, $\varepsilon_1(\omega)$ is the real part that aids in determining the polarization and dispersion of light, and $\varepsilon_2(\omega)$ is the imaginary part that describes the material's absorptive properties.Understanding the optical properties of this material, including refractive index $n(\omega)$,  absorption coefficient $\alpha(\omega)$, reflectivity $R(\omega)$, and energy loss spectra $L(\omega)$, is based on these components, $\varepsilon_1(\omega)$ and $\varepsilon_2(\omega)$. The following describes the relationship between the dielectric constants and other optical parameters:\cite{Saha2000,Fox2010,Hilal2016,Dar2019}

\begin{align}
    n(\omega) &= \frac{1}{\sqrt{2}} \left( \sqrt{\varepsilon_1^2(\omega) + \varepsilon_2^2(\omega)} + \varepsilon_1(\omega) \right)^{1/2} \\
    R(\omega) &= \left| \frac{\sqrt{\varepsilon(\omega)} - 1}{\sqrt{\varepsilon(\omega)} + 1} \right|^2 \\
    \alpha(\omega) &= \frac{2\omega k}{c} \\
    L(\omega) &= \frac{\varepsilon_2(\omega)}{\varepsilon_1^2(\omega) + \varepsilon_2^2(\omega)}
\end{align}
Specific electronic transitions between the valence and conduction bands are represented by the characteristic peaks that appear in the optical spectra\cite{Flores2008}.From Figure~\ref{fig:optical}(a) we can see that the static dielectric constant $\epsilon_1(0)$ is 2.66.The charge carrier recombination rate, which impacts optoelectronic device performance, is strongly influenced by the static dielectric constant\cite{Liu2018}.Materials with a high dielectric constant exhibit a lower charge carrier recombination rate, thereby enhancing device performance. It can also be seen from Figure~\ref{fig:optical} (a) that As energy increases, the dielectric function's real portion $\epsilon_1(\omega)$, begins to rise and reaches its maximum value of 7.01 at 5.57 eV.Then it starts to decrease and reaches to a minimum value of -1.64 at 7.27 eV.The negative value in the UV region means maximum reflectivity as electromagnetic waves gets reflected by medium in this energy range\cite{Hilal2016}.After that $\epsilon_1(\omega)$ starts to increase and approaches towards unity.

The dielectric function's imaginary portion $\epsilon_2(\omega)$, which describes the absorption characteristics of double perovskite materials and is inextricably tied to their electronic band structure, is shown in Figure~\ref{fig:optical}(b). The photon energy at which $\epsilon_2(\omega)$, first appears is approximately 3.52 eV, which is in good agreement with the band gap energies of the materials. This suggests that electronic transitions from the valence band to the conduction band occur as a result of the materials' absorption of photons at energies that match their band gaps.The prominent peak in $\epsilon_2(\omega)$ is observed at around 6.23 eV.The sharp decline after that suggests minimal interaction between the material's surface and the incident electromagnetic radiation\cite{Rouf2021}.At around 11.40 eV $\epsilon_2(\omega)$ approaches zero indicating that the material acts transparent above this energy range which is likely due to the absence of available electronic states for transition.

The absorption spectra reflects the distortion of light intensity as it traverses a unit distance in an absorbing medium. Figure~\ref{fig:optical}(c) illustrates the absorption spectra of Sr$_2$PrSbO$_6$ highlighting prominent peak between 6 eV to 11 eV covering the whole UV region; hence, the compound exhibits elevated absorbance and negligible electron losses in the ultraviolet area. Consequently, this double perovskite functions as effective UV absorbers, mitigating the detrimental impacts of UV radiation and presenting prospective uses in UV phototherapy and photovoltaic cells.Figure~\ref{fig:optical}(c) illustrates that the absorption coefficient $\alpha(\omega)$ is minimal across the visible spectrum and increases gradually in the ultraviolet region.The optical transparency in the visible spectrum indicates considerable potential for using these materials as a hole transport layer (HTM) in solar cells.

The refractive index $n(\omega)$ of Sr\textsubscript{2}PrSbO\textsubscript{6} closely resembles the shape of the real part of the dielectric function $\epsilon_1(\omega)$ as depicted in Figure~\ref{fig:optical} (e).The calculated static refractive index $\epsilon(0)$ is 1.63.The value of $n(\omega)$ is significantly high between 4 eV and 7.13 eV. It is important to emphasise that this material can function well as inner layer coatings in UV absorbing devices\cite{Dey2020} because of their high refractive index in the ultraviolet (UV) region. The refractive index then gradually drops and, at high energies, falls below unity, suggesting a weaker interaction between the material and incident electromagnetic radiation.

Electron energy loss spectra (EELS) characterise the energy dissipation of a high-velocity electron as it traverses a medium.
When electrons in the material absorb photon energy, a high-speed electron loses some of its energy, leading to collective electron oscillations that can be identified in the $L(\omega)$ spectra\cite{Kumar2020}. Figure~\ref{fig:optical}(f) presents  peak in the high eneergy region of the $L(\omega)$ spectra, signifying considerable electron losses in this area relative to the visible region. When the incident frequency of the electromagnetic wave is below the plasma frequency, electrons within the material can move sufficiently fast to produce plasma oscillations, which effectively screen or negate the external electric field. This leads to enhanced reflectivity, marked by peaks in the UV region of $L(\omega)$ spectra that align with minima in the reflectivity spectrum in Figure~\ref{fig:optical}(e).Sr$_2$PrSbO$_6$ exhibits a sharp energy loss peak at around 10.5 eV, corresponding to the plasmon resonance frequency. This suggests its suitability for plasmonic applications\cite{Ai2022}, such as sensors and plasmonic waveguides, where tunable optical properties are required.

In conclusion, the optical properties of Sr$_2$PrSbO$_6$ make this compound promising for widespread optoelectronic applications, ranging from high-speed photonics to solar energy harvesting and plasmonics.\cite{Ai2022,Yao2025}

\section{Thermoelectric Property}

Comprehending the thermoelectric characteristics of materials is crucial for transforming waste heat into usable energy. Annually, around 20-50\% of industrial energy intake is lost as waste heat\cite{SwRI}.Thermoelectric materials can directly transform thermal energy into electrical power, hence improving total energy efficiency. In this part, we have examined the thermoelectric characteristics of Sr$_2$PrSbO$_6$ within the temperature range of 300 to 700 K. The constant relaxation time approximation was employed to compute thermoelectric parameters including the Seebeck coefficient ($S$), electronic thermal conductivity per relaxation time ($\kappa_e / \tau_0$) and electrical conductivity per relaxation time ($\sigma / \tau_0$) using the BoltzTrap2 algorithm~\cite{Madsen2018}.

\begin{figure}[H]
    \centering

                    \includegraphics[width=0.7\linewidth]{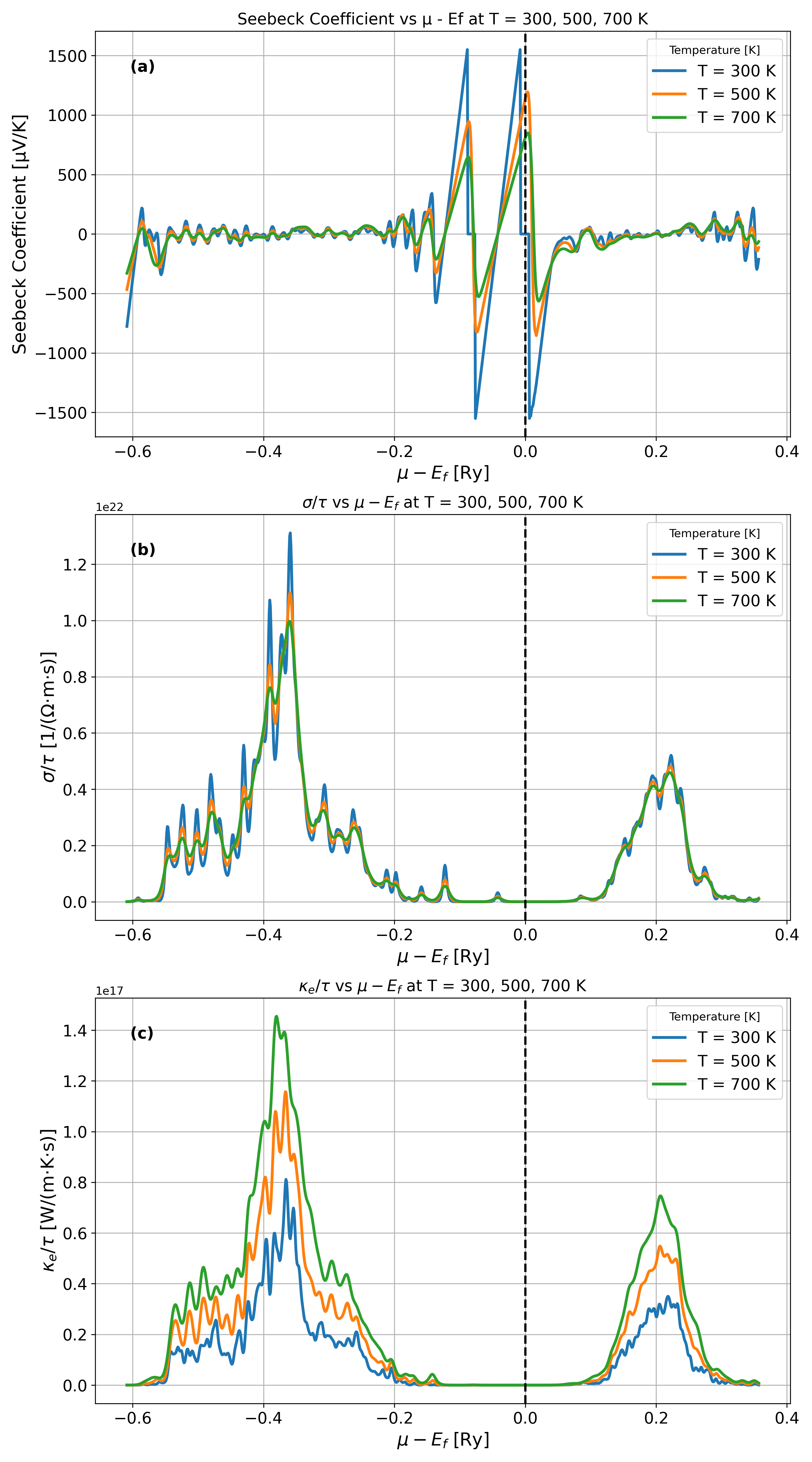}

        \caption{Variation in the thermoelectric properties of Sr\textsubscript{2}PrSbO\textsubscript{6} as a function of $\mu - E_f$ at different temperatures (300 K, 500 K, 700 K): (a) Seebeck coefficient (S); (b) Electrical conductivity scaled by relaxation time ($\sigma/\tau$); (c) Electronic thermal conductivity scaled by relaxation time ($\kappa_e/\tau$). The dashed vertical line indicates the Fermi level ($E_f$).}

    \label{fig:thermal}
\end{figure}

The Seebeck coefficient is measured in terms of voltage induced at the cost of temperature gradient across a conductor. The sign of $S$ denotes the nature of the dominant charge carriers. A positive $S$ indicates p-type semiconductors, where holes are the dominant charge carriers, whereas a negative $S$ denotes n-type semiconductors, characterized by electrons as the primary charge carriers. We simulated the Seebeck coefficient of Sr$_2$PrSbO$_6$ and plotted the relationship between the Seebeck coefficient $S$ and $\mu-E_f$ under three different temperature conditions (300 K, 500 K and 700 K), as shown in Figure~\ref{fig:thermal}(a).The data demonstrate that the compound exhibits substantial Seebeck coefficient values in both the positive and negative ranges, indicating the presence of both types of charge carriers in the systems. $S$ attains its greatest values in the range of $\mu-E_f$ values of -0.09 $R_y$ to 0.01 $R_y$ levels. The material demonstrates substantial Seebeck coefficients of approximately 1569 $\mu$K V$^{-1}$ at 300 K. When the temperature is increased to high temperature, the material under investigation show a decrease in $S$ value, indicating that at high temperature the Seebeck coefficient decreases.The compound would perform well as a thermoelectric material at around 300 K.

For optimal thermoelectric performance, a compound must demonstrate higher electrical conductivity. The plots of electrical conductivity versus $\mu-E_f$ at various temperatures are presented in Figure~\ref{fig:thermal}(b). It can be noticed from the plots that the value of $\sigma$ is higher at negative $\mu-E_f$ than in the positive $\mu-E_f$ region. This result implies that hole doping in this material will be more beneficial for thermoelectric applications than electron doping.
From the point of view of temperature change, the $\sigma$ values of Sr$_2$PrSbO$_6$ at low temperature contain many burr-like peaks, as shown in Figure~\ref{fig:thermal}(b). This indicates that the conductivity fluctuates greatly with the change in $\mu-E_f$, but the overall change trend is the same.

At 300 K, the maximum value of $\sigma$ for Sr$_2$PrSbO$_6$ is found to be approximately $1.1289 \times 10^{22}$ $\Omega^{-1}$m$^{-1}$s$^{-1}$. However, when the temperature is increased from 300 K to 900 K, $\sigma$ decreases with temperature. The slight decrease in the electrical conductivity with the increasing temperature may be attributed to increasing charge carrier concentration, collision, and scattering phenomena.
The phenomenon of heat conduction in a material by moment of free electrons and lattice vibration is termed thermal conductivity $\kappa$. This conductivity results from lattice vibrations occurring within the compounds during the motion of thermally excited electrons, encompassing both electronic $\kappa_e$ and phononic $\kappa_l$ components. This study calculates thermal conductivity while disregarding the phononic component, despite its contribution to the material's total thermal conductivity. This is because BoltzTraP2 code does not incorporate this contribution. This exclusion has been noticed in other comparable investigations owing to the identical computational limitation\cite{Khandy2017,Hasan2022}.Figure~\ref{fig:thermal}(c) illustrates the variation of electronic thermal conductivity versus $\mu-E_f$ at 300 K, 500 K, and 700 K.It can be noticed from the figure that both the compounds exhibited higher thermal conductivity for negative $\mu-E_f$, and their thermal conductivity is found to be $\sim 1.759\times10^17$ $W \Omega^{-1}$m$^{-1}$s$^{-1}$ at 300 K. However, in contrast to electrical conductivity, electronic thermal conductivity increases drastically when the temperature goes from 300 K to 700 K and achieves maximum peaks $\sim 1.759 \times 10^{17}$ W $\Omega^{-1}$m$^{-1}$s$^{-1}$. It can also be noticed from the plots that both electrical and thermal conductivities show similar profiles. Hence, the results follow the Wiedemann-Franz law.\cite{Yadav2019}

To enhance the precision of the assessment regarding the conversion of thermal to electrical energy, the figure of merit (ZT) was calculated near $E_f=0$ using the formula $ZT = \frac{\sigma S^2 T}{\kappa}$\cite{Olivieri2014} and summarised in Table~\ref{table:thermoelectric_properties}.The ZT value measures the highest conversion efficiency attainable by a thermoelectric device.It is noticeable that at 300K  the ZT value was found to be 0.331.
It is essential to recognize that the ZT values obtained have the potential for enhancement via the implementation of strain effects or the refinement of doping strategies\cite{Cheng-Wei2022}. These findings suggest that Sr$_2$PrSbO$_6$ is highly promising thermoelectric material, making it suitable for various applications in energy conversion technologies.

\begin{table}[H]
\centering
\caption{Thermoelectric Properties at Various Temperatures Near $E_f =0$}
\label{table:thermoelectric_properties}
\begin{tabular}{lllll}
\toprule
Temperature (K) & 300  & 500  & 700  \\
\midrule
$S$ ($\mu$V/K)  & 70 &   -18   & 11  \\
\(\sigma/\tau\) (1/\(\Omega\) m s) & \(5.4 \times 10^{21}\)  & \(3.9 \times 10^{21}\)  & \(3 \times 10^{21}\) \\
\(\kappa_e/\tau\) (W/mK s) & \(2.4 \times 10^{16}\)  & \(4.62 \times 10^{16}\)  & \(6.41 \times 10^{16}\) \\
ZT & 0.331  & 0.013  & 0.004 \\
\bottomrule
\end{tabular}
\end{table}

\section{Conclusion}

This study utilises first-principles Density Functional Theory (DFT) computations to examine the structural, electronic, optical, and thermoelectric properties of Sr\textsubscript{2}PrSbO\textsubscript{6}, a double perovskite with considerable potential for advanced engineering applications. The material crystallises in a cubic perovskite structure exhibiting high thermodynamic stability, as evidenced by its negative formation energy of $-22.77$ eV per formula unit. The calculated Goldschmidt tolerance factor of $0.968$ signifies structural stability, making it a suitable option for practical applications.

The investigation of the electronic band structure indicates a substantial direct bandgap of $3.488$ eV, categorising Sr\textsubscript{2}PrSbO\textsubscript{6} as a wide bandgap semiconductor which is appropriate for use in UV light-emitting diodes (LEDs), photovoltaic cells, and high-power electronics. Assessments of optical properties indicate significant absorption in the UV-visible spectrum and elevated refractive indices, reinforcing its applicability in optoelectronic devices and plasmonic applications. The thermoelectric investigation reveals a significant Seebeck coefficient and advantageous electrical and thermal conductivity, positioning Sr\textsubscript{2}PrSbO\textsubscript{6} as a viable choice for energy conversion technologies, including waste heat recovery systems.

The lack of magnetism, along with the material's resilience to breakdown and strong covalent bonding in B--O interactions, underscores its potential for high-temperature and chemically rigorous applications, such as electrocatalysis. These findings establish Sr\textsubscript{2}PrSbO\textsubscript{6} as a multifunctional material with potential to contribute significantly to next-generation electronic, photonic, and energy systems, making it a compelling subject for further experimental validation and engineering development.

This study establishes a foundation for utilising Sr\textsubscript{2}PrSbO\textsubscript{6} in industrial and technical applications, providing a significant contribution to the progression of functional materials in optoelectronics, thermoelectrics, and sustainable energy solutions.
\section{Declaration of Competing Interest}
The authors declare that they have no known competing financial interests or personal relationships that could have appeared to influence the work reported in this paper.

\section{Data Availability}
The data supporting the findings of this study are available upon request from the corresponding author. 
\section{Author Contributions}
Md. Mohiuddin designed the study and performed the calculations analyzed the data and wrote the manuscript. A. Kabir administered the project and supervised the investigations by guiding the computations, analyzed the data. All authors reviewed the manuscript.

\section{Acknowledgments}
The authors acknowledge the University of Dhaka for facilitating the research environment in the Department of Physics, and Bangladesh Research and Education Network BdREN (bdren.net.bd) for the computational lab facilities.


\begin{thebibliography}{99}
% Reference 1
\bibitem{Walter2010}
Walter, M. G. et al. Solar Water Splitting Cells. \textit{Chem. Rev.} \textbf{110}, 6446–6473 (2010).

% Reference 2
\bibitem{Zhang2015}
Zhang, J., Zhao, Z., Xia, Z. \& Dai, L. A metal-free bifunctional electrocatalyst for oxygen reduction and oxygen evolution reactions. \textit{Nat. Nanotechnol.} \textbf{10}, 444–452 (2015).

% Reference 3
\bibitem{Chen2011}
Chen, Y. W. et al. Atomic layer-deposited tunnel oxide stabilizes silicon photoanodes for water oxidation. \textit{Nat. Mater.} \textbf{10}, 539–544 (2011).

% Reference 4
\bibitem{Armand2008}
Armand, M. \& Tarascon, J.-M. Building better batteries. \textit{Nature} \textbf{451}, 652–657 (2008).

% Reference 5
\bibitem{Electrochemical1972}
Fujishima, A. \& Honda, K. Electrochemical photolysis of water at a semiconductor electrode. \textit{Nature} \textbf{238}, 37–38 (1972).

% Reference 6
\bibitem{Kudo2008}
Kudo, A. \& Miseki, Y. Heterogeneous photocatalyst materials for water splitting. \textit{Chem. Soc. Rev.} \textbf{38}, 253–278 (2008).

% Reference 7
\bibitem{Gasteiger2009}
Gasteiger, H. A. \& Marković, N. M. Just a Dream—or Future Reality? \textit{Science} \textbf{324}, 48–49 (2009).

% Reference 8
\bibitem{Bruce2012}
Bruce, P. G., Freunberger, S. A., Hardwick, L. J. \& Tarascon, J.-M. Li–O$_2$ and Li–S batteries with high energy storage. \textit{Nat. Mater.} \textbf{11}, 19–29 (2012).

% Reference 9
\bibitem{Zhu2017}
Zhu, S. \& Wang, D. Photocatalysis: Basic Principles, Diverse Forms of Implementations and Emerging Scientific Opportunities. \textit{Adv. Energy Mater.} \textbf{7}, 1700841 (2017).

% Reference 10
\bibitem{Boettcher2011}
Boettcher, S. W. et al. Photoelectrochemical Hydrogen Evolution Using Si Microwire Arrays. \textit{J. Am. Chem. Soc.} \textbf{133}, 1216–1219 (2011).

% Reference 11
\bibitem{Weng2017}
Weng, B. et al. A layered Na$_{1-x}$Ni$_y$Fe$_{1-y}$O$_2$ double oxide oxygen evolution reaction electrocatalyst for highly efficient water-splitting. \textit{Energy Environ. Sci.} \textbf{10}, 121–128 (2017).

% Reference 12
\bibitem{Bandgap2018}
Sun, S. et al. Bandgap Engineering of Stable Lead-Free Oxide Double Perovskites for Photovoltaics. \textit{Adv. Mater.} \textbf{30}, 1705901 (2018).

% Reference 13
\bibitem{Weng2018}
Weng, B. et al. Barium Bismuth Niobate Double Perovskite/Tungsten Oxide Nanosheet Photoanode for High-Performance Photoelectrochemical Water Splitting. \textit{Adv. Energy Mater.} \textbf{8}, 1701655 (2018).

% Reference 14
\bibitem{Perez2018}
Pérez-Tomás, A., Mingorance, A., Tanenbaum, D. \& Lira-Cantú, M. Metal Oxides in Photovoltaics: All-Oxide, Ferroic, and Perovskite Solar Cells. In \textit{The Future of Semiconductor Oxides in Next-Generation Solar Cells}, Elsevier, 267–356 (2018).

% Reference 15
\bibitem{Grinberg2013}
Grinberg, I. et al. Perovskite oxides for visible-light-absorbing ferroelectric and photovoltaic materials. \textit{Nature} \textbf{503}, 509–512 (2013).

% Reference 16
\bibitem{Ge2018}
Ge, J., Yin, W.-J. \& Yan, Y. Solution-Processed Nb-Substituted BaBiO$_3$ Double Perovskite Thin Films for Photoelectrochemical Water Reduction. \textit{Chem. Mater.} \textbf{30}, 1017–1031 (2018).

% Reference 17
\bibitem{Pena2001}
Peña, M. A. \& Fierro, J. L. G. Chemical Structures and Performance of Perovskite Oxides. \textit{Chem. Rev.} \textbf{101}, 1981–2018 (2001).

% Reference 18
\bibitem{Cohen1992}
Cohen, R. E. Origin of ferroelectricity in perovskite oxides. \textit{Nature} \textbf{358}, 136–138 (1992).

% Reference 19
\bibitem{García2012}
García-Martín, S., King, G., Nénert, G., Ritter, C. \& Woodward, P. M. The Incommensurately Modulated Structures of the Perovskites NaCeMnWO$_6$ and NaPrMnWO$_6$. \textit{Inorg. Chem.} \textbf{51}, 4007–4014 (2012).
% Reference 20

\bibitem{FernandezMartinez2012}
Fernández-Martínez, F., Montero, J. L., Carrillo, I. \& Colón, C. FT-Raman and FT-IR vibrational spectroscopic studies of Sr$_2$RESbO$_6$ (RE = La to Lu and Y) double perovskites. \textit{J. Alloys Compd.} \textbf{538}, 34–39 (2012). 


% Reference 20+1
\bibitem{Chen2019}
Chen, X., Xu, J., Xu, Y., Luo, F. \& Du, Y. Rare earth double perovskites: a fertile soil in the field of perovskite oxides. \textit{Inorg. Chem. Front.} \textbf{6}, 2226–2238 (2019).

% Reference 21
\bibitem{Rani2024}
Rani, M. et al. Rare earth-based oxides double perovskites A$_2$NiMnO$_6$ (A= La and Gd): Applications in magneto-caloric, photo-catalytic and thermoelectric devices. \textit{Phys. B Condens. Matter} \textbf{680}, 415645 (2024).

% Reference 22
\bibitem{Otsuka2015}
Otsuka, S. \& Hinatsu, Y. Structures and magnetic properties of rare earth double perovskites containing antimony or bismuth Ba$_2$LnMO$_6$ (Ln=rare earths; M=Sb, Bi). \textit{J. Solid State Chem.} \textbf{227}, 132–141 (2015).

% Reference 23
\bibitem{Mahato2016}
Mahato, D. K., Rudra, M. \& Sinha, T. P. Structural and electrical features of rare earth based double perovskite oxide: Pr$_2$NiZrO$_6$. \textit{J. Alloys Compd.} \textbf{689}, 617–624 (2016).

% Reference 24
\bibitem{Das2022}
Das, P. et al. Investigation of electrical and magnetic properties of rare-earth based double perovskite: Dy$_2$CoMnO$_6$. \textit{Ferroelectrics} \textbf{588}, 1–8 (2022).

% Reference 25
\bibitem{Dar2023}
Dar, S. A. et al. Study of structural, electronic, magnetic, and optical properties of A$_2$FeMnO$_6$ (A = Ba, La) double perovskites, experimental and DFT analysis. \textit{Colloids Surf. A Physicochem. Eng. Asp.} \textbf{664}, 131145 (2023).

% Reference 26
\bibitem{Kumar2022}
Kumar, A. et al. Experimental and theoretical studies of structural, electronic and magnetic properties of RE$_2$NiCrO$_6$ (RE= Ce, Pr and Nd) double perovskites. \textit{Physica B: Condensed Matter} \textbf{633}, 413801 (2022).

% Reference 27
\bibitem{Laghzaoui2023}
Laghzaoui, S., Lamrani, A. F. \& Laamara, R. A. Robust half-metallic ferromagnet in doped double perovskite Sr$_2$TiCoO$_6$ by rare-earth elements for photovoltaic and thermoelectric conversion: A DFT method. \textit{J. Phys. Chem. Solids} \textbf{183}, 111639 (2023).

% Reference 28
\bibitem{AlQaisi2023}
Al-Qaisi, S. et al. A comprehensive first-principles study on the physical properties of Sr$_2$ScBiO$_6$ for low-cost energy technologies. \textit{Opt. Quantum Electron.} \textbf{55}, 1015 (2023).

% Reference 29
\bibitem{Hanif2022}
Hanif, M. et al. Theoretical investigation of physical properties of Sr$_2$XNbO$_6$ (X = La, Lu) double perovskite oxides for optoelectronic and thermoelectric applications. \textit{Int. J. Energy Res.} \textbf{46}, 16884–16902 (2022).

% Reference 30
\bibitem{Haid2019}
Haid, S. et al. Thermoelectric, Structural, Optoelectronic and Magnetic properties of double perovskite Sr$_2$CrTaO$_6$: First principle Study. \textit{Mater. Sci. Eng. B} \textbf{245}, 68–74 (2019).

% Reference 31
\bibitem{Sun2017}
Sun, Q. \& Yin, W.-J. Thermodynamic Stability Trend of Cubic Perovskites. \textit{J. Am. Chem. Soc.} \textbf{139}, 14905–14908 (2017).

% Reference 32
\bibitem{Hudgins2003}
Hudgins, J. L., Simin, G. S., Santi, E. \& Khan, M. A. An assessment of wide bandgap semiconductors for power devices. \textit{IEEE Trans. Power Electron.} \textbf{18}, 907–914 (2003).

% Reference 33
\bibitem{Hohenberg1964}
Hohenberg, P. \& Kohn, W. Inhomogeneous Electron Gas. \textit{Phys. Rev.} \textbf{136}, B864–B871 (1964).

% Reference 34
\bibitem{Payne1992}
Payne, M. C., Teter, M. P., Allan, D. C., Arias, T. A. \& Joannopoulos, J. D. Iterative minimization techniques for ab initio total-energy calculations: molecular dynamics and conjugate gradients. \textit{Rev. Mod. Phys.} \textbf{64}, 1045–1097 (1992).

% Reference 35
\bibitem{Kohn1965}
Kohn, W. \& Sham, L. J. Self-Consistent Equations Including Exchange and Correlation Effects. \textit{Phys. Rev.} \textbf{140}, A1133–A1138 (1965).

% Reference 36
\bibitem{Gillan1989}
Gillan, M. J. Calculation of the vacancy formation energy in aluminium. \textit{J. Phys. Condens. Matter} \textbf{1}, 689 (1989).

% Reference 37
\bibitem{Vanderbilt1990}
Vanderbilt, D. Soft self-consistent pseudopotentials in a generalized eigenvalue formalism. \textit{Phys. Rev. B} \textbf{41}, 7892–7895 (1990).

% Reference 38
\bibitem{Kresse1996}
Kresse, G. \& Furthmüller, J. Efficiency of ab-initio total energy calculations for metals and semiconductors using a plane-wave basis set. \textit{Comput. Mater. Sci.} \textbf{6}, 15–50 (1996).

% Reference 39
\bibitem{Kresse1999}
Kresse, G. \& Joubert, D. From ultrasoft pseudopotentials to the projector augmented-wave method. \textit{Phys. Rev. B} \textbf{59}, 1758–1775 (1999).

% Reference 40
\bibitem{Paier2005}
Paier, J., Hirschl, R., Marsman, M. \& Kresse, G. The Perdew-Burke-Ernzerhof exchange-correlation functional applied to the G2-1 test set using a plane-wave basis set. \textit{J. Chem. Phys.} \textbf{122}, 234102 (2005).

% Reference 41
\bibitem{Perdew1996}
Perdew, J. P., Burke, K. \& Ernzerhof, M. Generalized Gradient Approximation Made Simple. \textit{Phys. Rev. Lett.} \textbf{77}, 3865–3868 (1996).

% Reference 42
\bibitem{Wang2021}
Wang, V., Xu, N., Liu, J.-C., Tang, G. \& Geng, W.-T. VASPKIT: A user-friendly interface facilitating high-throughput computing and analysis using VASP code. \textit{Comput. Phys. Commun.} \textbf{267}, 108033 (2021).

% Reference 43
\bibitem{Madsen2018}
Madsen, G. K. H., Carrete, J. \& Verstraete, M. J. BoltzTraP2, a program for interpolating band structures and calculating semi-classical transport coefficients. \textit{Comput. Phys. Commun.} \textbf{231}, 140–145 (2018).

% Reference 44
\bibitem{Sato2016}
Sato, T., Takagi, S., Deledda, S., Hauback, B. C. \& Orimo, S. Extending the applicability of the Goldschmidt tolerance factor to arbitrary ionic compounds. \textit{Sci. Rep.} \textbf{6}, 23592 (2016).

% Reference 45
\bibitem{Jin2021}
Jin, S. Can We Find the Perfect A-Cations for Halide Perovskites? \textit{ACS Energy Lett.} \textbf{6}, 3386–3389 (2021).

% Reference 46
\bibitem{Melius1992}
Melius, C. F. The Thermochemistry and Reaction Pathways of Energetic Material Decomposition and Combustion. \textit{Philos. Trans. R. Soc. Lond., Phys. Sci. Eng.} \textbf{339}(1654), 365–376 (1992).

% Reference 47
\bibitem{Volonakis2016}
Volonakis, G., Filip, M. R., Haghighirad, A. A., Sakai, N., Wenger, B., Snaith, H. J. \& Giustino, F. Lead-Free Halide Double Perovskites via Heterovalent Substitution of Noble Metals. \textit{J. Phys. Chem. Lett.} \textbf{7}, 1254–1259 (2016).

% Reference 48
\bibitem{Charifi2023}
Charifi, Z., Baaziz, H., Uğur, Ş. \& Uğur, G. Prediction of the electronic structure, optical and vibrational properties of ScXCo$_2$Sb$_2$ (X = V, Nb and Ta) double half-Heusler alloys: a theoretical study. \textit{Indian J. Phys.} \textbf{97}, 413–428 (2023).

% Reference 49
\bibitem{Abdullah2023}
Abdullah, D. \& Gupta, D. C. Exploring the half-metallic ferromagnetism, dynamical and mechanical stability, optoelectronic and thermoelectric properties of K$_2$NaMI$_6$ (M = Mn, Co, Ni) for spintronic applications. \textit{Sci. Rep.} \textbf{13}, 1–16 (2023).

% Reference 50
\bibitem{Touati2006}
Touati, F., Mnif, F. \& Lawati, A. High-Temperature Electronics: Status and Future Prospects in the 21st Century. \textit{J. Eng. Res. TJER} \textbf{3}, 43–54 (2006).

% Reference 51
\bibitem{Chen2024}
Chen, Q., Zhou, L., Zhang, J., Chen, D., Zhu, W., Xi, H., Zhang, J., Zhang, C. \& Hao, Y. Recent Progress of Wide Bandgap Perovskites towards Two-Terminal Perovskite/Silicon Tandem Solar Cells. \textit{Nanomaterials} \textbf{14}, 202 (2024).

% Reference 52
\bibitem{Forgacs2017}
Forgács, D., Pérez-del-Rey, D., Ávila, J., Momblona, C., Gil-Escrig, L., Dänekamp, B., Sessolo, M. \& Bolink, H. J. Efficient wide band gap double cation – double halide perovskite solar cells. \textit{J. Mater. Chem. A} \textbf{5}(7), 3203–3207 (2017).

% Reference 53
\bibitem{Fang2024}
Fang, Z., Deng, B., Jin, Y., Yang, L., Chen, L., Zhong, Y., Feng, H., Yin, Y., Liu, K., Li, Y., Zhang, J., Huang, J., Zeng, Q., Wang, H., Yang, X., Yang, J., Tian, C., Xie, L., Wei, Z., \& Xu, X. Surface reconstruction of wide-bandgap perovskites enables efficient perovskite/silicon tandem solar cells. \textit{Nat. Commun.} \textbf{15}(1), 10554 (2024).

% Reference 54
\bibitem{Shahzad2024}
Shahzad, M. K., Hussain, S., Riaz, M., Sattar, H., Ashraf, G. A., Azeem, W., Ali, S. M. \& Alam, M. Investigation of ultra wide bandgap Flouro-perovskite materials RBeF$_3$ (R = K and Li) for smart window applications: A DFT study. \textit{Heliyon} \textbf{10}(7), e29143 (2024).

% Reference 55
\bibitem{Zhou2023}
Zhou, X., Lu, Z., Zhang, L. \& Ke, Q. Wide-bandgap all-inorganic lead-free perovskites for ultraviolet photodetectors. \textit{Nano Energy} \textbf{117}, 108908 (2023).

% Reference 56
\bibitem{Grimaud2013}
Grimaud, A., May, K. J., Carlton, C. E., Lee, Y.-L., Risch, M., Hong, W. T., Zhou, J. \& Shao-Horn, Y. Double perovskites as a family of highly active catalysts for oxygen evolution in alkaline solution. \textit{Nat. Commun.} \textbf{4}, 2439 (2013).

% Reference 57
\bibitem{SpringerLink2022}
Solid State Properties: From Bulk to Nano. \url{https://link.springer.com/book/10.1007/978-3-662-55922-2} (SpringerLink, 2022).

% Reference 58
\bibitem{Fox2010}
Fox, M. \textit{Optical Properties of Solids}, Vol. 3. Oxford University Press (2010).

% Reference 59
\bibitem{Albanesi2005}
Albanesi, E. A., Peltzer y Blanca, E. L. \& Petukhov, A. G. Calculated optical spectra of IV–VI semiconductors PbS, PbSe and PbTe. \textit{Comput. Mater. Sci.} \textbf{32}, 85–95 (2005).

% Reference 60
\bibitem{Kumar2020}
Kumar, V. \& Roy, D. R. Strain-induced band modulation and excellent stability, transport and optical properties of penta-MP$_2$ (M = Ni, Pd, and Pt) monolayers. \textit{Nanoscale Adv.} \textbf{2}(10), 4566–4580 (2020).

% Reference 61
\bibitem{Saha2000}
Saha, S., Sinha, T. P. \& Mookerjee, A. Electronic structure, chemical bonding, and optical properties of paraelectric BaTiO$_3$. \textit{Phys. Rev. B} \textbf{62}, 8828–8834 (2000).

% Reference 62
\bibitem{Dar2019}
Dar, S. A., Sharma, R., Srivastava, V. \& Sakalle, U. K. Investigation on the electronic structure, optical, elastic, mechanical, thermodynamic and thermoelectric properties of wide band gap semiconductor double perovskite Ba$_2$InTaO$_6$. \textit{RSC Adv.} \textbf{9}, 9522–9532 (2019).

% Reference 63
\bibitem{Hilal2016}
Hilal, M., Rashid, B., Khan, S. H. \& Khan, A. Investigation of electro-optical properties of InSb under the influence of spin-orbit interaction at room temperature. \textit{Mater. Chem. Phys.} \textbf{184}, 41–48 (2016).

% Reference 64
\bibitem{Flores2008}
Flores, M. Z. S., Freire, V. N., dos Santos, R. P., Farias, G. A., Caetano, E. W. S., de Oliveira, M. C. F., Fernandez, J. R. L., Scolfaro, L. M. R., Bezerra, M. J. B., Oliveira, T. M., Bezerra, G. A., Cavada, B. S. \& Leite Alves, H. W. Optical absorption and electronic band structure first-principles calculations of $\alpha$-glycine crystals. \textit{Phys. Rev. B} \textbf{77}, 115104 (2008).

% Reference 65
\bibitem{Liu2018}
Liu, X., Xie, B., Duan, C., Wang, Z., Fan, B., Zhang, K., Lin, B., Colberts, F. J. M., Ma, W., Janssen, R. A. J., Huang, F. \& Cao, Y. A high dielectric constant non-fullerene acceptor for efficient bulk-heterojunction organic solar cells. \textit{J. Mater. Chem. A} \textbf{6}, 395–403 (2018).

% Reference 66
\bibitem{Rouf2021}
Rouf, S. A., Hussain, M. I., Mumtaz, U., Majeed, A. M. \& Masood, H. T. A density functional theory study of the structural, electronic, and optical properties of XGaO$_3$ (X = V, Nb) perovskites for optoelectronic applications. \textit{J. Comput. Electron.} \textbf{20}(4), 1484–1495 (2021).

% Reference 67
\bibitem{Dey2020}
Dey, A., Baraiya, B. A., Adhikary, S. \& Jha, P. K. First-principles calculations of the effects of edge functionalization and size on the band gap of Be$_3$N$_2$ nanoribbons: Implications for nanoelectronic devices. \textit{ACS Appl. Nano Mater.} \textbf{4}(1), 493–502 (2020).

% Reference 68
\bibitem{Ai2022}
Ai, B., Fan, Z. \& Wong, Z. J. Plasmonic–perovskite solar cells, light emitters, and sensors. \textit{Microsyst. Nanoeng.} \textbf{8}(1), 5 (2022).

% Reference 69
\bibitem{Yao2025}
Yao, Y., Li, B., Ding, D., Kan, C., Hang, P., Zhang, D., Hu, Z., Ni, Z., Yu, X., \& Yang, D. Oriented wide-bandgap perovskites for monolithic silicon-based tandems with over 1000 hours operational stability. \textit{Nat. Commun.} \textbf{16}(1), 40 (2025).

% Reference 70
\bibitem{SwRI}
Southwest Research Institute. Waste heat recovery research.

% Reference 71
\bibitem{Khandy2017}
Khandy, S. A. \& Gupta, D. C. Investigation of structural, magneto-electronic, and thermoelectric response of ductile SnAlO$_3$ from high-throughput DFT calculations. \textit{Int. J. Quantum Chem.} \textbf{117}(8), e25351 (2017).

% Reference 72
\bibitem{Hasan2022}
Hasan, S. \textit{et al.} First-principles calculations of thermoelectric transport properties of quaternary and ternary bulk chalcogenide crystals. \textit{Materials} \textbf{15}(8), 2843 (2022).

% Reference 73
\bibitem{Yadav2019}
Yadav, A., Deshmukh, P. C., Roberts, K., Jisrawi, N. M. \& Valluri, S. R. An analytic study of the Wiedemann–Franz law and the thermoelectric figure of merit. \textit{J. Phys. Commun.} \textbf{3}(10), 105001 (2019).

% Reference 74
\bibitem{Olivieri2014}
Olivieri, A. C. \& Escandar, G. M. Analytical Figures of Merit. In \textit{Practical Three-Way Calibration}, edited by A. C. Olivieri \& G. M. Escandar, pp. 93–107. Elsevier, Boston, 2014.

% Reference 75
\bibitem{Cheng-Wei2022}
Cheng-Wei, W. \textit{et al.} Enhanced high-temperature thermoelectric performance by strain engineering in BiOCl. \textit{Phys. Rev. Appl.} \textbf{18}(1), 014053 (2022).

\end{thebibliography}
\end{document}